\documentclass{emulateapj}
\usepackage{apjfonts}
\usepackage{xspace}
\usepackage{amsmath}
\usepackage{mathtools}
\usepackage{hyperref}
\usepackage{epsf}
\usepackage{epstopdf}

\hypersetup{%
    colorlinks=false, 
    urlcolor=magenta, 
    linkcolor=blue, 
    citecolor=blue,
    breaklinks=true, 
    bookmarksnumbered=true,
    bookmarksopen=false, 
    bookmarksopenlevel=0,
    hypertexnames=true, 
    unicode=true,          % non-Latin characters in Acrobat’s bookmarks
    pdftoolbar=true,        % show Acrobat’s toolbar?
    pdfmenubar=true,        % show Acrobat’s menu?
    pdffitwindow=false,     % window fit to page when opened
    pdfstartview={FitH}    % fits the width of the page to the window
}

\usepackage{etoolbox}
\makeatletter

\patchcmd{\NAT@citex}
  {\@citea\NAT@hyper@{\NAT@nmfmt{\NAT@nm}\NAT@date}}
  {\@citea\NAT@nmfmt{\NAT@nm}\NAT@hyper@{\NAT@date}}
  {}% Do nothing if patch works
  {}% Do nothing if patch fails

\patchcmd{\NAT@citex}
  {\@citea\NAT@hyper@{%
     \NAT@nmfmt{\NAT@nm}%
     \hyper@natlinkbreak{\NAT@aysep\NAT@spacechar}{\@citeb\@extra@b@citeb}%
     \NAT@date}}
  {\@citea\NAT@nmfmt{\NAT@nm}%
   \NAT@aysep\NAT@spacechar%
   \NAT@hyper@{\NAT@date}}
  {}% Do nothing if patch works
  {}% Do nothing if patch fails

\patchcmd{\NAT@citex}
  {\@citea\NAT@hyper@{%
     \NAT@nmfmt{\NAT@nm}%
     \hyper@natlinkbreak{\NAT@spacechar\NAT@@open\if*#1*\else#1\NAT@spacechar\fi}%
       {\@citeb\@extra@b@citeb}%
     \NAT@date}}
  {\@citea\NAT@nmfmt{\NAT@nm}%
   \NAT@spacechar\NAT@@open\if*#1*\else#1\NAT@spacechar\fi%
   \NAT@hyper@{\NAT@date}}
  {}% Do nothing if patch works
  {}% Do nothing if patch fails

\makeatother

\begin{document}
\shorttitle{Computing Hydrostatic Mass Bias}
\shortauthors{Lau, Nagai, \& Nelson}
\slugcomment{The Astrophysical Journal, accepted}
\journalinfo{The Astrophysical Journal, accepted{\rm }}
%\submitted{Received; accepted; published}

\title{Weighing Galaxy Clusters with Gas. I. \\ On the Methods of Computing Hydrostatic Mass Bias}

\author{Erwin T. Lau\altaffilmark{1,2}}
\author{Daisuke Nagai\altaffilmark{1,2,3}}
\author{Kaylea Nelson\altaffilmark{3}}

%\altaffiltext{1}{Department of Physics, Yale University, New Haven, CT 06520, USA; \email{erwin.lau@yale.edu}} 
%\altaffiltext{2}{Yale Center for Astronomy and Astrophysics, Yale University, New Haven, CT 06520, USA} 
%\altaffiltext{3}{Department of Astronomy, Yale University, New Haven, CT 06520, USA} 

\affil{ 
$^1$Department of Physics, Yale University, New Haven, CT 06520, USA; \href{mailto:erwin.lau@yale.edu}{erwin.lau@yale.edu} \\
$^2$Yale Center for Astronomy and Astrophysics, Yale University, New Haven, CT 06520, USA \\
$^3$Department of Astronomy, Yale University, New Haven, CT 06520, USA
}


\begin{abstract}
Mass estimates of galaxy clusters from X-ray and Sunyeav-Zel'dovich observations assume the intracluster gas is in hydrostatic equilibrium with their gravitational potential. However, since galaxy clusters are dynamically active objects whose dynamical states can deviate significantly from the equilibrium configuration, the departure from the hydrostatic equilibrium assumption is one of the largest sources of systematic uncertainties in cluster cosmology. In the literature there has been two methods for computing the hydrostatic mass bias based on the Euler and the modified Jeans equations, respectively, and there has been some confusion about the validity of these two methods.  The word ``Jeans'' was a misnomer, which incorrectly implies that the gas is collisionless.  To avoid further confusion, we instead refer these methods as ``summation'' and ``averaging'' methods respectively. In this work, we show that these two methods for computing the hydrostatic mass bias are equivalent by demonstrating that the equation used in the second method can be derived from taking spatial averages of the Euler equation. Specifically, we identify the correspondences of individual terms in these two methods mathematically and show that these correspondences are valid to within a few percent level using hydrodynamical simulations of galaxy cluster formation. In addition, we compute the mass bias associated with the acceleration of gas and show that its contribution is small in the virialized regions in the interior of galaxy clusters, but becomes non-negligible in the outskirts of massive galaxy clusters. We discuss future prospects of understanding and characterizing biases in the mass estimate of galaxy clusters using both hydrodynamical simulations and observations and their implications for cluster cosmology. 
\end{abstract}

%-------------------------------------------------%
\section{Introduction}
%-------------------------------------------------%

Clusters of galaxies are the largest virialized objects in the Universe, promising to provide unique insights into both cosmology and astrophysics. Mass of galaxy clusters is one of the fundamental quantities for using clusters as cosmological and astrophysical probes. Cluster mass estimates from X-ray and Sunyeav-Zel'dovich (SZ) data assume the hydrostatic equilibrium (HSE) of gas in their gravitational potential. However, galaxy clusters are dynamical active objects, whose dynamical state deviates significantly from the equilibrium configuration. As a result, the hydrostatic mass bias is one of the largest sources of systematic uncertainties in cluster cosmology \citep{allen_etal08, vikhlinin_etal09, planck_XX13}. To exploit the statistical power of the ongoing and upcoming multi-wavelength cluster surveys, the hydrostatic mass bias has to be understood and controlled at the level of a few percent. 

To date, numerical simulations have been used extensively to characterize the hydrostatic mass bias and understand its origin.  Hydrodynamical cluster simulations predict that the hydrostatic bias is at the level of $5\%$ to $35\%$, depending on the dynamical state of clusters \citep[e.g.,][]{rasia_etal06,nagai_etal07b, jeltema_etal08, piffaretti_etal08, burns_etal10, meneghetti_etal10}.  These simulations suggest that the bias is predominantly due to the effective non-thermal pressure support provided by bulk and turbulent gas flows induced by cluster mergers and accretion events \citep{lau_etal09, vazza_etal09, nelson_etal12}. Comparison of weak lensing and X-ray hydrostatic mass further suggests that the hydrostatic mass bias at the level of $\lesssim 10\%$ for the relaxed clusters and $15\%$ to $20\%$ for dynamically active systems \cite[e.g.,][]{zhang_etal10, mahdavi_etal13}. 

In the literature, there are two methods for computing the bias. In the first method, the mass enclosed within a surface is determined by summing up contribution of each gas element on the surface its potential gradient, which is evaluated from the thermal pressure gradient and temporal and spatial gradients of gas velocities using the Euler equation \citep{fang_etal09, suto_etal13}. The second method, on the other hand, uses the potential gradient averaged over the surface, which is estimated from the averaged gas densities, pressure, and velocities using a modified version of the Jeans equation \citep{rasia_etal04, lau_etal09, nelson_etal12}.  There has been a question regarding the validity of the second method \citep{suto_etal13}. This confusion has partly stemmed from the use of the word ``Jeans'', a misnomer, which incorrectly implies that the gas is collisionless.  To avoid further confusion, we refer them as the ``summation'' and ``averaging'' methods respectively hereafter. 

Different authors have made a number of simplifying assumptions in computing the corrections terms in the hydrostatic mass bias with both methods. In \citet{rasia_etal04} the mean and random components of the gas motions are not differentiated. In \citet{fang_etal09} and \citet{lau_etal09}, the support due to rotational motions and streaming motions are explicitly included, but each arrived at a different conclusion about the relative importance of rotational versus random motions partly because of the different physical meanings of mass correction terms in the summation and averaging methods used respectively in those work. More recently \citet{suto_etal13} relaxed the steady state assumption of the cluster (i.e., $\partial {\bf v}/\partial t=0$), and suggested that the acceleration of gas introduces extra bias whose magnitude is comparable to other mass bias terms. 

The primary goal of this work is to assess the validity of two methods used to compute the hydrostatic mass bias and to understand its physical origin. In this work, we show that the summation and averaging methods for computing the hydrostatic mass bias are both valid by demonstrating that the averaging method can be derived from the summation method by applying spatial averaging over gas elements. This process introduces additional mean and dispersion terms in gas velocities that are absent in the original Euler equation. These extra terms originate from fluctuations in gas velocities that are implicitly included in the summation method, but must be explicitly accounted for in the averaging method.  Using hydrodynamical simulations of cluster formation, we show that the correspondence between these two methods are robust at the level of a few percent.  In addition, we compute the acceleration term directly using multiple time-steps of each simulated cluster and assess its relative importance. Finally, we argue that the averaging method is more suitable than the summation method for application to observational datasets. 

This paper is organized as follows. In Section~\ref{sec:theory} we present the theoretical frameworks of cluster mass reconstruction using the summation and averaging methods. In Section~\ref{sec:sim}, using numerical simulations we show the equivalence of the two methods and evaluate the importance of gas acceleration. In Section~\ref{sec:summary} we provide a summary and discuss implications of our results.   

%-------------------------------------------------%
\section{Mass Reconstruction: Theory}
\label{sec:theory}
%-------------------------------------------------%

%-------------------------------------------------%
\subsection{Summation Method}
\label{sec:sum}
%-------------------------------------------------%

Using Gauss's law for the gravitational field, the total gravitational mass enclosed within volume $V$ with surface $\partial V$ is 
\begin{equation}
M = \frac{1}{4\pi G}\oint_{\partial V} {\bf \nabla} \Phi {\bf \cdot} d{\bf S} \, ,
\label{eqn:gauss}\\
\end{equation}
where $M$ is the enclosed mass and $\Phi$ is the gravitational potential. The mass inside this surface is known when the potential gradient ${\bf \nabla} \Phi$ is known at every position on the imaginary surface with differential surface element $d{\bf S}$. 

The potential gradient is generally given by the dynamical evolution of any single particle component that constitutes the system. We start with the mass and momentum conservation equations in index notation: 
\begin{align}
\frac{\partial \rho}{\partial t} + \frac{\partial \left(\rho \bar{v}^{i}\right) }{\partial x^i} &= 0, \label{eqn:con}\\
\frac{\partial \left(\rho \bar{v}^{i}\right)}{\partial t} + \frac{\partial \tau^{ij} }{\partial x^j} &= -\rho g^{ij}\frac{\partial \Phi}{\partial x^j}, \label{eqn:mom}
\end{align}
where $\rho$ is the particle density, $v^i$ is the $i$-th component of the particle velocity, and $g^{ij}$ is the spatial metric tensor. Repeated indices are summed over. The momentum flux tensor (or stress tensor) is defined as 
\begin{equation}
\tau^{ij}  \equiv  \rho\overline{v^iv^j} = \rho\sigma^{2,ij}+\rho u^iu^j \, ,
\label{eqn:mometum_flux_tensor}
\end{equation}
where $u^i = \bar{v}^i$ and $\sigma^{2,ij}\equiv \overline{\left(v^i-u^i\right)\left(v^j-u^j\right)}$ is the velocity dispersion tensor. The overline denotes averaging over some volume of the system. 

In the hydrodynamical limit where the mean free path of the gas particle is small compared to the scale of the system, the gas particles undergo frequent collisions and their distribution is approximately Maxewellian. For such gas, viscosity is negligible and $\sigma^{2,ij}$ is isotropic with zero off-diagonal components. The momentum flux tensor is then given by
\begin{equation}
\tau^{ij} = \tau_{E}^{ij} \equiv Pg^{ij}+\rho u^iu^j \, ,
\label{eqn:euler_mft}
\end{equation}
where $P$ is the thermal pressure. The momentum conservation equation (Equation~\ref{eqn:mom}) then becomes the Euler equation when combined with the continuity equation (Equation~\ref{eqn:con}): 
\begin{equation}
\frac{\partial u^{i}}{\partial t} + u^j\frac{\partial u^{i} }{\partial x^j} = -\frac{1}{\rho}\frac{\partial P}{\partial x_i} -\frac{\partial \Phi}{\partial x_i}.
\label{eqn:euler}
\end{equation}
Using Gauss's Law (Equation~\ref{eqn:gauss}), the mass is given by 
\begin{equation}
M= \frac{-1}{4\pi G}\oint_{\partial V} \left(\frac{\partial u^i}{\partial t}+u^j\frac{\partial u^i}{\partial x^j}+\frac{1}{\rho}\frac{\partial P}{\partial x_i}\right) dS_i .
\end{equation}
This mass can be broken down into various effective mass terms \citep{fang_etal09, suto_etal13}: 
\begin{equation}
M(<r) = M^{S}_{\rm tot}(<r) = M^{S}_{\rm therm} + M^{S}_{\rm rot} + M^{S}_{\rm stream} +M^{S}_{\rm accel},
\end{equation}
where the superscript $S$ indicates that the mass terms are derived from applying the summation method. The individual terms in spherical coordinates $(r,\theta,\phi)$ are
\begin{align}
&M^{S}_{\rm therm} = \frac{-1}{4\pi G}\int \frac{1}{\rho}\frac{\partial P}{\partial r} r^2 d\Omega ,\label{eqn:euler_therm} \\
&M^{S}_{\rm rot} = \frac{1}{4\pi G}\int \left( u_{\theta}^2+u_{\phi}^2 \right) r\,d\Omega , \label{eqn:euler_rot}\\
&M^{S}_{\rm stream} = \frac{-1}{4\pi G}\int \left(u_r\frac{\partial u_r}{\partial r} + \frac{u_\theta}{r}\frac{\partial u_r }{\partial \theta} + \frac{u_\phi}{r\sin \theta}\frac{\partial u_r}{\partial \phi}\right) r^2 d\Omega,  \label{eqn:euler_stream}\\
&M^{S}_{\rm accel} =  \frac{-1}{4\pi G}\int \frac{\partial u_r}{\partial t} r^2 d\Omega,
\label{eqn:euler_accel}
\end{align}
where $d\Omega = \sin\theta d\theta d\phi$ is the solid angle element and we have adopted $\partial V$ to be a spherical surface with radius $r$. The physical significance of the terms are as follows: 
$M^{S}_{\rm therm}$ is the term representing the support against gravity from the thermal pressure of the gas; 
$M^{S}_{\rm rot}$ is the sum of contribution of support due to tangential gas motions (which includes both mean and random motions); 
$M^{S}_{\rm stream}$ is the sum of support due to spatial variations of streaming gas motions in the radial direction; 
and $M^{S}_{\rm accel}$ is the sum of support due to temporal variations in the radial gas velocities, which is negative (positive) if there is net gas acceleration (deceleration) from the cluster center. 
%{\bf Note that the term ``acceleration'' does not mean that we are following the trajectory of the gas element, since we are using the Eulerian time derivative $\partial/\partial t$: the acceleration term merely tracks the change in gas velocity at a fixed location with respect to time. }

%-------------------------------------------------%
\subsection{Averaging Method}
\label{sec:ave}
%-------------------------------------------------%

In the hydrodynamical limit, each gas element follows the Euler equation (Equation~\ref{eqn:euler}).  The momentum flux tensor for each {\em single} gas element is $\tau^{ij}=\tau_{E}^{ij}$, where $u^i$ is the $i$-th component of the mean velocity of the  gas element. Let $\langle \cdots \rangle_w$ denote the averaging operator performed over some volume $V$ containing a number of gas elements. The average value of any arbitrary quantity $f\,({\bf x})$ is given by
\begin{equation}
\langle f \rangle_w({\bf x}) = \int_V f \left({\bf x}-{\bf x}'\right) w\left({\bf x};{\bf x}'\right) d^3x', 
\end{equation}
where $w$ is some normalized weight function $\int_V w\, d^3x=1$. When $w$ is independent of ${\bf x}$, as in the case of volume averaging, $\langle f \rangle_w({\bf x})$ is essentially the convolution $f \ast w$. The averaging operator $\langle \cdots \rangle_w$ then commutes with partial differentiation with respect to $x$, which follows from the convolution theorem: $\partial_x \langle f \rangle_w = \partial_x \left(f \ast w\right) = \partial_x f \ast w = \langle \partial_x f \rangle_w$. 
For density-weighted averaging, the weight function is the normalized density distribution, and the averaging operator can be expressed as $\langle \cdots \rangle_\rho = \langle \rho \cdots \rangle/\langle \rho \rangle$. Applying this volume averaging to the Euler momentum flux tensor $\tau_{E}^{ij}$ (Equation~\ref{eqn:euler_mft}), we have
\begin{align}
\langle \tau_{E}^{ij} \rangle &= \langle Pg^{ij} \rangle + \langle \rho u^iu^j \rangle \nonumber \\ 
&= \langle Pg^{ij}\rangle  + \langle \rho \rangle \langle u^iu^j \rangle_\rho \nonumber \\
&= \langle Pg^{ij}\rangle  + \langle \rho \rangle \sigma_{\rho}^{2\,ij} + \langle \rho \rangle \langle u^i\rangle_\rho\langle u^j \rangle_\rho \, ,
\label{eqn:avgmom}
\end{align}
where $\langle u^i \rangle_\rho$ is the $i$-th component density-weighted average velocity of the gas elements in the considered volume $V$, and $\sigma_{\rho}^{2,ij} \equiv \left\langle \left(u^i-\langle u^i\rangle_\rho \right)\left(u^j-\langle u^j\rangle_\rho \right)\right\rangle_\rho$ is the density-weighted velocity dispersion tensor in the same volume $V$, which in general can be anisotropic and contain non-zero off-diagonal components.

Putting this averaged Euler momentum flux tensor (Equation~\ref{eqn:avgmom}) into the momentum conservation equation (Equation~\ref{eqn:mom}) and averaging other terms (e.g., $\rho u^i \rightarrow \langle \rho u^i \rangle = \langle\rho\rangle \langle u^i \rangle_\rho$) results in the following equation:
\begin{align}
\frac{\partial \langle u^{i}\rangle_\rho}{\partial t} + \langle u^j\rangle_\rho\frac{\partial \langle u^{i}\rangle_\rho }{\partial x^j} &+ \frac{1}{\langle\rho\rangle} \frac{\partial \langle \rho \rangle \sigma_{\rho}^{2,ij}}{\partial x^i} \nonumber \\
&= -\frac{1}{\langle\rho\rangle}\frac{\partial \langle P \rangle}{\partial x_i} -\frac{1}{\langle\rho\rangle}\left\langle \rho\frac{\partial \Phi}{\partial x_i}\right\rangle. 
\label{eqn:jeans}
\end{align}
Here we have used the fact that the averaging operator $\langle \cdots \rangle$ commutes with the partial differential operators $\partial_t$ and $\partial_x$. Note that while this equation formally resembles the Jeans equation plus a thermal pressure gradient term, it does not imply that the gas is collisionless. This process of spatial averaging of the Euler equation is analogous to spatially filtering the Navier-Stokes equation that yields the Reynolds equation, in the presence of viscosity \citep[e.g.,][]{tennekes_lumley72}. 

In spherical coordinates, the radial component of Equation~(\ref{eqn:jeans}) is given by
\begin{align}
&\frac{\partial \langle u_{r}\rangle_\rho}{\partial t} + \langle u_r\rangle_\rho\frac{\partial \langle u_r\rangle_\rho }{\partial r} + \frac{\langle u_\theta\rangle_\rho}{r}\frac{\partial \langle u_r\rangle_\rho }{\partial \theta}  + \frac{\langle u_\phi\rangle_\rho}{r\sin \theta}\frac{\partial \langle u_r\rangle_\rho }{\partial \phi} \nonumber \\ 
&{}+ \frac{1}{\langle\rho\rangle}\left(\frac{\partial \langle \rho \rangle \sigma_{\rho,rr}^{2}}{\partial r} + \frac{1}{r}\frac{\partial \langle \rho \rangle \sigma_{\rho,r\theta}^{2}}{\partial \theta} + \frac{1}{r\sin\theta}\frac{\partial \langle \rho \rangle \sigma_{\rho,r\phi}^{2}}{\partial \phi}\right) \nonumber \\
&{}+\frac{1}{r}\left(2\sigma_{\rho,rr}^2-\sigma_{\rho,\theta\theta}^2-\sigma_{\rho,\phi\phi}^2-\langle u_\theta\rangle_\rho^2-\langle u_\phi\rangle_\rho^2+\sigma_{\rho,r\theta}^2\cot\theta\right) \nonumber \\
&= -\frac{1}{\langle\rho\rangle}\frac{\partial \langle P \rangle}{\partial r} -\frac{1}{\langle\rho\rangle}\left\langle\rho\frac{\partial \Phi}{\partial r}\right\rangle.
\label{eqn:jeans_radial}
\end{align}

Assuming spherical symmetry and averaging over the surface of the imaginary sphere with radius $r$, the last term in this equation can be expressed in terms of the mass enclosed within $r$: 
\begin{equation}
\frac{1}{\langle\rho\rangle}\left\langle\rho\frac{\partial \Phi}{\partial r}\right\rangle = \frac{1}{\langle\rho\rangle} \left\langle \rho \frac{GM(<r)}{r^2} \right\rangle = \frac{GM(<r)}{r^2} 
\end{equation}
where we have used the fact that mass is unchanged by averaging.  As with the summation method, the total mass enclosed within radius $r$ can also be broken down into different effective mass terms:
\begin{align}
M(<r) = M^{A}_{\rm tot}(<r) &= M^{A}_{\rm therm} + M^{A}_{\rm rand} + M^{A}_{\rm rot} \nonumber \\ 
&+ M^{A}_{\rm cross}+ M^{A}_{\rm stream} +M^{A}_{\rm accel},
\end{align}
with
\begin{align}
&M^{A}_{\rm therm} = \frac{-r^2}{G\langle\rho\rangle}\frac{\partial \langle P \rangle}{\partial r}, \label{eqn:jeans_therm}\\
&M^{A}_{\rm rand} = \frac{-r^2}{G\langle\rho\rangle}\frac{\partial \langle \rho \rangle \sigma_{\rho,rr}^{2}}{\partial r} - \frac{r}{G}\left(2\sigma_{\rho,rr}^2-\sigma_{\rho,\theta\theta}^2-\sigma_{\rho,\phi\phi}^2\right), \label{eqn:jeans_rand}\\ 
&M^{A}_{\rm rot} = \frac{r}{G}\left(\langle u_\theta\rangle_\rho^2+\langle u_\phi\rangle_\rho^2\right),  \label{eqn:jeans_rot}\\
&M^{A}_{\rm stream} = \frac{-r^2}{G}\left(\langle u_r\rangle_\rho\frac{\partial \langle u_r\rangle_\rho }{\partial r} + \frac{\langle u_\theta\rangle_\rho}{r}\frac{\partial \langle u_r\rangle_\rho }{\partial \theta} +\frac{\langle u_\phi\rangle_\rho}{r\sin \theta}\frac{\partial \langle u_r\rangle_\rho }{\partial \phi}\right),  \label{eqn:jeans_stream}\\
&M^{A}_{\rm cross} = \frac{-r^2}{G\langle\rho\rangle}\left(\frac{1}{r}\frac{\partial \langle \rho \rangle \sigma_{\rho,r\theta}^{2}}{\partial \theta} + \frac{1}{r\sin\theta}\frac{\partial \langle \rho \rangle \sigma_{\rho,r\phi}^{2}}{\partial \phi}\right) \nonumber \\
&\qquad \quad {}-\frac{r}{G}\left( \sigma_{\rho,r\theta}^2\cot\theta \right),  \label{eqn:jeans_cross}\\
&M^{A}_{\rm accel} =  \frac{-r^2}{G}\frac{\partial \langle u_{r}\rangle_\rho}{\partial t}, \label{eqn:jeans_accel}
\end{align}
where the superscript $A$ denotes that the mass terms are derived from the averaging method (Equation~\ref{eqn:jeans}). Except for $M^{A}_{\rm accel}$, all of the above mass terms are the same as Equations~(7) -- (11) in \citet{lau_etal09} where the gas density and pressure are volume averaged and the velocities are density-weighted averaged over the spherical surface. The physical significance of the terms are as follows: 
$M^{A}_{\rm therm}$ is the term representing the support against gravity from the averaged thermal pressure of the gas;
$M^{A}_{\rm rand}$ is the support from the random motions of gas in both the radial and tangential directions;
$M^{A}_{\rm rot}$ is the rotational support due to {\em mean} tangential motions of gas;
$M^{A}_{\rm stream}$ comes from spatial variations of the {\em mean} radial streaming gas velocities; 
$M^{A}_{\rm cross}$ arises from the off-diagonal components of the velocity dispersion tensor, which are non-zero if the radial and tangential components of the random motions are correlated;
and $M^{A}_{\rm accel}$ is the support due to to temporal variations of the {\em mean} radial gas velocities at a fixed radius, which is negative (positive) for net gas accelerating (decelerating) away from the cluster center. 

%-------------------------------------------------%
\subsection{Correspondence between the Summation and Averaging Methods}
\label{sec:interp}
%-------------------------------------------------%

Both $M^{S}_{\rm tot}$ and $M^{A}_{\rm tot}$ are estimates of the total mass using gas properties. They are both derived from the Euler equations with the application of Gauss's Law, but there are notable differences in how gas properties are handled. In the summation method, gas properties are summed over the surface containing the enclosed mass. In the averaging method, the quantities are averaged over the spherical surface. Specifically, each mass term in $M^{S}_{\rm tot}$ has corresponding term in $M^{A}_{\rm tot}$,
\begin{align}
M^{S}_{\rm therm}  &\Leftrightarrow  M^{A}_{\rm therm},   \\
M^{S}_{\rm stream} &\Leftrightarrow  M^{A'}_{\rm rand}  + M^{A}_{\rm stream} + M^{A}_{\rm cross},  \\
M^{S}_{\rm rot}  &\Leftrightarrow  M^{A'}_{\rm rot},  \\
M^{S}_{\rm accel}  &\Leftrightarrow M^{A}_{\rm accel}, 
\end{align}
where 
\begin{align}
&M^{A'}_{\rm rand}  = M^{A}_{\rm rand} - \frac{r}{G}\left(\sigma_{\rho,\theta\theta}^2+\sigma_{\rho,\phi\phi}^2\right),\\
&M^{A'}_{\rm rot} = M^{A}_{\rm rot} + \frac{r}{G}\left(\sigma_{\rho,\theta\theta}^2+\sigma_{\rho,\phi\phi}^2\right). 
\label{eqn:MJ_primed}
\end{align}
Note that the rotation term in the summation method is different from the one in the averaging method: $M^S_{\rm rot}$ includes both the {\em mean} and {\em random} parts of the tangential components, while $M^A_{\rm rot}$ only accounts for the {\em mean} tangential motions. When comparing the rotational term between the two methods, one should use $M^{A'}_{\rm rot}$ which includes both the {\em mean} and {\em random} parts of the tangential components. Likewise for the streaming term, $M^S_{\rm stream}$ should be compared to $M^{A'}_{\rm stream}$ instead of $M^{A}_{\rm stream}$. In the following section, we use numerical simulations to demonstrate the equivalence of the two methods by evaluating individual terms in $M^{S}_{\rm tot}$ and $M^{A}_{\rm tot}$. 

%-------------------------------------------------%
\section{Mass Reconstruction: Simulation}
\label{sec:sim}
%-------------------------------------------------%

%-------------------------------------------------%
\subsection{Data}
\label{sec:data}
%-------------------------------------------------%

In this work, we use high resolution cosmological hydrodynamical simulations presented in \citet{nagai_etal07a, nagai_etal07b, lau_etal09, nelson_etal12}. Here we provide a brief description of the simulations, and we refer the readers to \citet{nagai_etal07a, nagai_etal07b} for details. The simulations are based on the flat concordance $\Lambda$CDM model: $\Omega_{\rm m}=1-\Omega_{\Lambda}=0.3$, $\Omega_{\rm b}=0.04286$, $h=0.7$ and $\sigma_8=0.9$, where the Hubble constant is defined as $100h{\ \rm km\ s^{-1}\ Mpc^{-1}}$, and $\sigma_8$ is the power spectrum normalization on an $8h^{-1}$~Mpc scale.  They were performed with the Adaptive Refinement Tree (ART) $N$-body$+$gasdynamics code \citep{kra99,kra02,rudd_etal08}, an Eulerian code that uses adaptive refinement in space and time, and (non-adaptive) refinement in mass \citep{klypin_etal01} to reach the high dynamic range required to resolve cores of halos formed in self-consistent cosmological simulations. They were run using a uniform 128$^3$ grid and 8 levels of mesh refinement in computational boxes of $120\,h^{-1}$~Mpc and $80\,h^{-1}$~Mpc on a side with peak resolution of $\approx 3.66\,h^{-1}$~kpc and $2.44\,h^{-1}$~kpc respectively.  The dark matter (DM) particle mass in the regions around each cluster was $9.1\times 10^{8}\,h^{-1}\, {M_{\odot}}$  and $2.7\times 10^{8}\,h^{-1}\,{M_{\odot}}$ for the two box sizes, while other regions were simulated with lower mass resolution. 

For this paper, we analyzed five most relaxed clusters at $z=0.026$ in the sample, re-simulated with non-radiative gas physics used in both \citet{lau_etal09} and \citet{nelson_etal12}. We selected relaxed clusters to focus on a clean sample ideal for testing the two methods in question. We examine the dependence of hydrostatic mass bias on cluster dynamical states in our second paper \citep{nelson_etal13}. Table~\ref{tab:sim} gives their $r_{500c}$, which is the radius enclosing an average total mass density of 500 times the critical density of the universe, and the corresponding enclosed mass $M_{500c}$. As shown in \citet{lau_etal09}, radiative cooling, star formation and supernova feedback have little effect on the hydrostatic mass bias and total mass recovery from gas motions, although the relative contributions of different mass terms do change near the cluster core. 

%%%%%%%%%%%%%%%%%%%%%%%%%%%%%%%%%%%%%%%%%%%%%%%%%%%%%
\begin{table}[t]
\begin{center}
\caption{Properties of the simulated clusters}\label{tab:sim}
\begin{tabular}{l | c c c  }
\hline
\hline
Cluster ID\hspace*{5mm}
&{$M_{500c}$} 
&{$r_{500c}$} 
& Box size \\
& {[10$^{14}$ $h^{-1} M_{\odot}$]}
& {[$h^{-1}$ Mpc]} 
& {[$h^{-1}$ Mpc]} \\
\hline
CL104 & 5.19 & 0.956 & 120\\
CL3 & 2.09 & 0.706 & 80\\
CL7 & 1.18 & 0.584 & 80\\
CL10 & 0.62 & 0.471 & 80\\
CL14 & 0.64 & 0.576 & 80\\
\hline
\end{tabular}
\end{center}
\end{table}
%%%%%%%%%%%%%%%%%%%%%%%%%%%%%%%%%%%%%%%%%%%%%%%%%%%%

%-------------------------------------------------%
\subsection{Method}
\label{sec:method}
%-------------------------------------------------%

To compute each mass term in both the summation and averaging methods presented in Section~\ref{sec:theory}, we work in the spherical coordinate system $(r,\theta,\phi)$, and divide the analysis region into 80 spherical logarithmic bins from $10\,h^{-1}{\rm kpc}$ to $10 \,h^{-1}{\rm Mpc}$ in the radial direction from the cluster center, defined as the position with the maximum binding energy. Each spherical bin is further subdivided into 60 and 120 uniform angular bins in the $\theta$ and $\phi$ directions, respectively. Our results are insensitive to the exact choice of binning. We choose the rest frame of the system to be the center-of-mass velocity of the total mass interior to each radial bin. This is different from \citet{lau_etal09} where the rest frame of the system is defined to be the mass-weighted average DM velocity interior to $r_{500c}$, but the results presented here are insensitive to the choice of the rest frame. We rotate the coordinate system for each radial bin such that the $z$-axis aligns with the axis of the total gas angular momentum of that bin. 

For $M^S_{\rm tot}$, we compute gas velocities, density and pressure in each angular bin by taking average values (with appropriate weighting) of each quantity for the hydro cells residing in the angular bin. The derivatives of velocity and pressure are computed directly by differencing the neighboring angular bins. We then sum the relevant terms times the surface area $r^2\Delta\cos\theta\Delta\phi$ for each angular bin over the spherical surface $4\pi r^2$ as in Equations~(\ref{eqn:euler_therm})~--~(\ref{eqn:euler_stream}). For $M^A_{\rm tot}$, we compute each term in Equations~(\ref{eqn:jeans_therm})~--~(\ref{eqn:jeans_stream}) by averaging values of the angular bins over the radial bin. The acceleration term $M^S_{\rm accel} = M^A_{\rm accel}$ is computed explicitly by taking the difference of the radial velocity at the same radial bin between two consecutive time-steps divided by the time between these two snapshots. One cluster in our sample (CL104) has fine time resolution of $\sim 0.04$ Gyr, while the other clusters have coarser time resolution of $\sim 0.35$ Gyr. We have checked that the acceleration term is insensitive to the choice of time resolution within this range. We remove large gas substructures that may bias the global gas pressure and velocity gradients by applying the clump exclusion method presented in \citet{zhuravleva_etal13}. In addition, we smooth each mass term by applying the Savitzky-Golay filter used in \citet{lau_etal09}. Finally, the true mass $M_{\rm true}$ is directly computed from the simulations. 

%-------------------------------------------------%
\subsection{Results}
\label{sec:results}
%-------------------------------------------------%

%%%%%%%%%%%%%%%%%%%%%%%%%%%%%%%%%%%%%%%%%%%%%%%%
\begin{figure}[t]
\begin{center}
\includegraphics[scale=0.7]{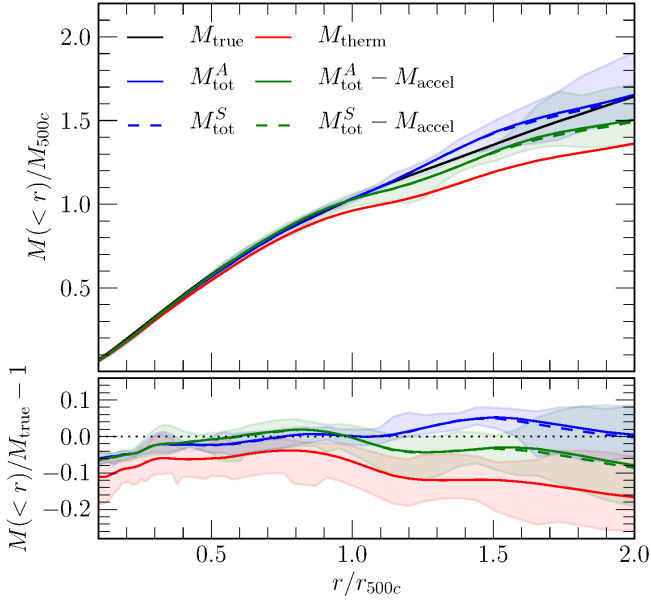}
\caption
{{\em Upper} panel: comparison between the true mass $M_{\rm true}$ (black line), the hydrostatic mass $M_{\rm therm}$ (red line) and total recovered masses $M^S_{\rm tot}$ (dashed lines) and $M^A_{\rm tot}$ (solid lines) with acceleration terms (blue lines) and without (green lines).  The profiles are averaged over the cluster sample, with mass and radius normalized to $M_{500c}$ and $r_{500c}$. The shaded regions show $\pm1\sigma$ scatter. {\em Lower} panel: comparison between the deviation from the true mass $M_{\rm true}$ for the hydrostatic mass  $M_{\rm therm}$ (red line) and recovered mass $M^S$ (dashed lines) and $M^A$ (solid lines) with acceleration terms (blue lines) and without (green lines).  The shaded regions show $\pm1\sigma$ scatter for the terms in the averaging method (solid lines). }
\label{fig:f1}
\end{center}
\end{figure}
%%%%%%%%%%%%%%%%%%%%%%%%%%%%%%%%%%%%%%%%%%%%%%%%

%%%%%%%%%%%%%%%%%%%%%%%%%%%%%%%%%%%%%%%%%%%%%%%%
\begin{figure}[t]
\begin{center}
\includegraphics[scale=0.7]{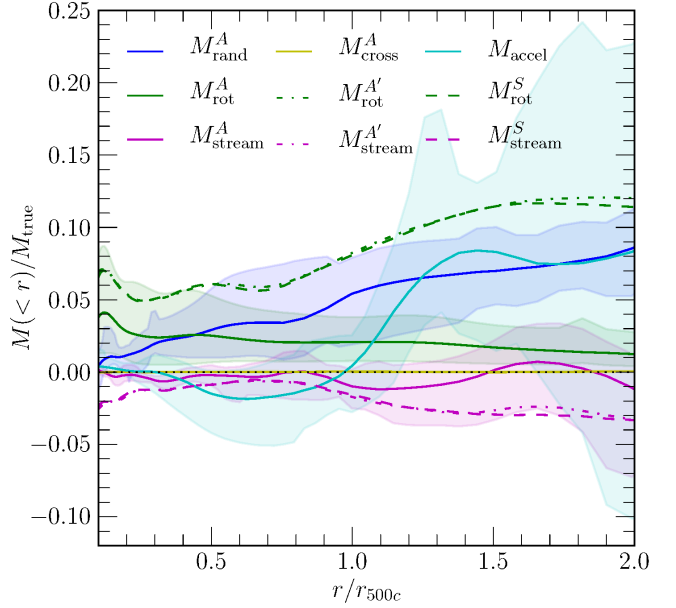}
\caption
{Comparison between the the different mass terms divided by the true mass in the mass estimates based on the summation method: $M^A_{\rm rot}$ (green dashed), $M^A_{\rm stream}$ (magenta dashed); and the averaging method $M^A_{\rm rand}$ (blue solid), $M^A_{\rm rot}$ (green solid), $M^A_{\rm stream}$ (magenta solid), $M^A_{\rm cross}$ (yellow solid). The dot-dashed lines are $M^{A'}_{\rm rot}$ (green) and $M^{A'}_{\rm stream}$, which correspond to $M^S_{\rm rot}$ and $M^S_{\rm stream}$  The profiles are averaged over the cluster sample, with mass and radius normalized to $M_{500c}$ and $r_{500c}$.  The shaded regions show $\pm1\sigma$ scatter for the terms in the averaging method (solid lines). }
\label{fig:f2}
\end{center}
\end{figure}
%%%%%%%%%%%%%%%%%%%%%%%%%%%%%%%%%%%%%%%%%%%%%%%%

Figure~\ref{fig:f1} shows the comparison between the true mass $M_{\rm true}$, hydrostatic mass $M^S_{\rm therm}$ and $M^A_{\rm therm}$, and the recovered masses $M^S_{\rm tot}$ and $M^A_{\rm tot}$, together with their deviations from the true mass as a function of normalized cluster centric radius, $r/r_{500c}$. We note the the hydrostatic mass for both methods are essentially identical $M_{\rm therm} = M^S_{\rm therm} = M^A_{\rm therm}$. On average, the recovery of the true mass profile is accurate at the level of a few percent at $r \approx r_{500c}$ for both summation and averaging methods. The acceleration term makes small contribution at $r \lesssim r_{500c}$, but its contribution becomes non-negligible at $r>r_{500c}$. For $r > r_{500c}$, $M^S_{\rm tot}$ and $M^A_{\rm tot}$ overestimates the true mass by $\sim 5\%$, mainly due to the acceleration term $M_{\rm accel}$. Ignoring this term leads to a slight underestimate of the true mass.

Figure~\ref{fig:f2} shows the comparison of different mass terms in $M^S$ and $M^A$ in units of the true mass. For the averaging method, the random motion $M^A_{\rm rand}$ is a dominant term, contributing to $\sim 5\%$ of the mass support at $r=r_{500c}$, followed by the rotational term $M^A_{\rm rot}$, the acceleration term $M_{\rm accel}$, and the streaming term $M^A_{\rm stream}$. The crossing term $M^A_{\rm cross}$ is very small ($\sim 0.01\%$), suggesting that the correlation among different velocity components is close to zero. At $r \gtrsim r_{500c}$, while the random motion term $M^A_{\rm rand}$ increases with radius, the acceleration term becomes comparable to $M^A_{\rm rand}$, but with large fluctuations. The streaming term $M^A_{\rm stream}$ is small compared to $M^A_{\rm rand}$. This term and the crossing term $M^A_{\rm cross}$ are ignored in \citet{nelson_etal12} as they are negligibly small. However for individual clusters the streaming term can become comparable to the random motion term. For the Euler's case, the dominant term at all radii is the $M^S_{\rm rot}$, which represents the sum of support from tangential motions from all gas elements. The acceleration term again can be comparable to $M^S_{\rm rot}$, although with large scatter. The streaming term $M^S_{\rm stream}$ is small but non-negligible. 

Figure~\ref{fig:f2} also shows the comparison of the rotational and streaming terms for the summation and averaging methods: $M^{A'}_{\rm rot}$, and $M^{A'}_{\rm stream}$. In Section~\ref{sec:interp} we showed that the two are mathematically identical, and it is indeed the case for our simulated clusters: i.e., the terms in these two methods agree to better than a few percent for the relaxed clusters. Note that $M^S_{\rm rot} \simeq M^{A'}_{\rm rot} > M^{A}_{\rm rot}$ because both $M^S_{\rm rot}$ and $M^{A'}_{\rm rot}$ account for the {\em mean} and {\em random} components of the tangential motions, while  $M^{A}_{\rm rot}$ only accounts for the mean component. This partly explains the difference in the contribution from rotational support between \citet{fang_etal09} and \citet{lau_etal09} who computed $M^S_{\rm rot}$ and $M^A_{\rm rot}$, respectively, despite the fact that they analyzed the same set of simulated clusters. 

Our averaged $M^S_{\rm tot}$ agree qualitatively with the results of \citet{suto_etal13}. The acceleration terms $M_{\rm accel}$ in their work and ours become large in the cluster outskirts. Their streaming term $M^S_{\rm stream}$ is slightly larger in magnitude, but still consistent within $\sim 2\sigma$ scatter of our results. Their rotational term $M^S_{\rm rot}$ is also slightly larger than ours, mainly because their simulated cluster includes effects of radiative cooling and star formation that lead to ``overcooling",  thereby increasing the rotational motion in the cluster core \citep{fang_etal09, lau_etal11}. This effect is also sensitive to the actual implementation of gas physics and the dynamical state of the cluster. 

%-------------------------------------------------%
\section{Summary and Discussion}
\label{sec:summary}
%-------------------------------------------------%

In this work, we show that two methods of computing the hydrostatic mass bias based on summation and averaging are equally valid and in fact equivalent.  In the first method, contributions from individual gas element are summed over the imaginary surface one by one. The second method, on the other hand, uses gas velocities averaged over the surface. We show that the averaging method can be derived from the summation method by spatially averaging the terms in Euler equation. Specifically, we identify the correspondences of individual terms in these two methods mathematically and show that they are indeed valid to within few percent using hydrodynamical simulations of galaxy cluster formation. In addition, we compute mass bias corrections associated with gas acceleration, which is generally small in the interior virialized regions of galaxy clusters, but becomes non-negligible in the outskirts of massive clusters where materials are still actively accreting today. 

In observations we generally do not have access to small-scale properties of the intracluster medium below the instrumental resolution limits. Therefore the averaging method is more suitable for analyzing observational datasets because it is based on spatially-averaged quantities that can be derived directly from observations with finite spatial resolution. For example, the mean gas velocity and gas velocity dispersion averaged over some range of spatial scales in the averaging method can be measured with the upcoming ASTRO-H mission\footnote{\url{http://astro-h.isas.jaxa.jp/}} via Doppler shift and broadening of heavy ion lines. Observationally, it is also difficult to measure gas acceleration, which introduces an irreducible bias. Simulations of cluster formation may be helpful in calibrating the gas acceleration term associated with the change in gas accretion rate and/or gas velocity profiles. 

There are additional limitations for both methods when analyzing real clusters. In this work, we have assumed that the cluster gas is inviscid and unmagnetized and its mean free path is much smaller than the spatial resolution of the simulations. However, the mean free path of the gas in the cluster outskirts can become large such that the hydrodynamic approximation breaks down. In real clusters where the intracluster medium is likely to have tangled magnetic fields (as the magnetic energy density is considerably smaller than the thermal energy density), electron-ion pairs in the intracluster plasma perform random walk, giving rise to a finite effective mean free path, which might be a finite fraction of the Spitzer's value. Finite mean free path also gives a rise to non-zero physical viscosity and transport coefficients, whose exact values are still unknown.  A non-zero viscous stress tensor can then introduce additional support against gravity, and it must be included in the momentum flux tensor in both summing and averaging methods. Magnetic field and cosmic rays further give rise to additional mass correction terms associated with magnetic and cosmic ray pressure as well as their advection terms at the level of few percent.  Even when the magnetic field is dynamically unimportant, magnetothermal instability may amplify turbulent motions, which could provide additional $5\%$ to $30\%$ of non-thermal pressure support \citep{parrish_etal12}. All these effects can change the relative contributions of different types of gas motions to the effective mass correction terms. 

In this work, we focused on a small number of relaxed clusters in order to test two methods in questions. However, mass biases are expected to be larger for unrelaxed clusters.  It is therefore important to characterize the distribution of hydrostatic mass biases for a wide range of masses, redshifts, dynamical states, and accretion histories by analyzing a large cosmologically representative sample of galaxy clusters.  In our second paper \citep{nelson_etal13}, we use a large statistical mass-limited sample of simulated galaxy clusters to investigate these issues. Such work might also help better assess the current tension between Planck primary CMB and SZ cluster counts results \citep{planck_XX13}. 
\vspace{10 mm}
\acknowledgments
We thank Andrea Morandi, Elena Rasia, Suto et al., and the anonymous referee for useful comments on the manuscript. This work was supported in part by NSF grant AST-1009811, NASA ATP grant NNX11AE07G, NASA Chandra Theory grant GO213004B, the Research Corporation, and by the facilities and staff of the Yale University Faculty of Arts and Sciences High Performance Computing Center. 

\pdfbookmark{REFERENCES}{references}
\bibliographystyle{apj_ads}\bibliography{ms}

\end{document}